\documentclass[english,aps,prb,preprint,superscriptaddress,showpacs,showkeys]{revtex4-1}
\usepackage[T1]{fontenc}
\usepackage[latin9]{inputenc}
\usepackage{ifsym}
\usepackage{amsmath}
\usepackage{amssymb}
\usepackage{wasysym}
\usepackage{graphicx}

\makeatletter
 
 \@ifundefined{textcolor}{}
 {%
   \definecolor{BLACK}{gray}{0}
   \definecolor{WHITE}{gray}{1}
   \definecolor{RED}{rgb}{1,0,0}
   \definecolor{GREEN}{rgb}{0,1,0}
   \definecolor{BLUE}{rgb}{0,0,1}
   \definecolor{CYAN}{cmyk}{1,0,0,0}
   \definecolor{MAGENTA}{cmyk}{0,1,0,0}
   \definecolor{YELLOW}{cmyk}{0,0,1,0}
 }

\makeatletter

\usepackage{amsfonts}

\usepackage{times}

\usepackage{epstopdf}

\makeatletter

 \usepackage{epstopdf}
    \usepackage{times}
\makeatother

\makeatother

\usepackage{babel}
\begin{document}

\title{Step Free Energies at Faceted Solid-Liquid Interfaces from Equilibrium
Molecular Dynamics Simulations}

\author{T. Frolov and M. Asta}

\affiliation{Department of Materials Science and Engineering, University of California,
Berkeley, California 94720, USA}
\begin{abstract}
In this work a method is proposed for computing step free energies
for faceted solid-liquid interfaces based on atomistic simulations.
The method is demonstrated in an application to (111) interfaces in
elemental Si, modeled with the classical Stillinger-Weber potential.
The approach makes use of an adiabatic trapping procedure, and involves
simulations of systems with coexisting solid and liquid phases separated
by faceted interfaces containing islands with different sizes, for
which the corresponding equilibrium temperatures are computed. We
demonstrate that the calculated coexistence temperature is strongly
affected by the geometry of the interface. We find that island radius
is inversely proportional to superheating, allowing us to compute
the step free energy by fitting simulation data within the formalism
of classical nucleation theory. The step free energy value is computed
to be $\gamma_{st}=0.103\pm0.005\times10^{-10}$ J/m. The approach
outlined in this work paves the way to the calculation of step free
energies relevant to the solidification of faceted crystals from liquid
mixtures, as encountered in nanowire growth by the vapor-liquid-solid
mechanism and in alloy casting. The present work also shows that at
low undercoolings the Stillinger-Weber interatomic potential for Si
tends to crystallize in the wurtzite, rather than the diamond-cubic
structure.
\end{abstract}


\maketitle

\section{Introduction\label{sec:Introduction}}

The properties of solid-liquid interfaces are known to play critical
roles in governing defect formation and the evolution of microstructural
patterns during the solidification of a crystal from its melt.\cite{Asta09rev}
In applications of crystal-growth theories and mesoscale simulation
methods to the study of solidification phenomena, quantitative information
is often required for the thermodynamic and kinetic properties of
solid-liquid interfaces, and the degrees to which these properties
vary with crystal orientation. For materials such as metals where
crystal-melt interfaces are molecularly rough, the most relevant properties
in this context include the magnitude and crystalline anisotropy of
the interfacial stiffness and mobility,\cite{Asta09rev,Hoyt2003121}
which dictate, respectively, the capillary undercooling at a growing
crystal-melt interface, and the relationship between local undercooling
and interface velocity. For many crystalline compounds and covalent
solids, the degree of crystalline anisotropy in crystal-melt interfacial
free energies can be highly pronounced, such that the equilibrium
crystal Wulff shape is characterized by faceted orientations. In this
case, crystal-growth kinetics are typically strongly influenced by
the properties of steps at faceted solid-liquid interfaces.\cite{ChernovRev2004,Jackson58}
For example, the excess free energies and mobilities of these linear
defects play critical roles in governing the rate of nucleation and
the growth of crystal islands during layer-by-layer growth of faceted
crystals, as observed in the synthesis of nanowires by the vapor-liquid-solid
mechanism.\cite{wagner:89,ChernovRev2004,Tersoff2011,golovin:074301,ChernovMRS}

Due to the difficulty inherent in performing direct experimental measurements
of the thermodynamic and kinetic properties of crystal-melt interfaces,
numerous computational approaches have been developed for their calculation
within the framework of atomistic simulations.\cite{Asta09rev,Mishin2010}
Specifically, for crystal-melt interfaces in systems with molecularly-rough
interfaces, equilibrium\cite{Broughton86,Davidchak00,Hoyt01,Horbach2012,Horbach11epl,Frolov09jcp,Frolov10p2,Frolov11a,Morris02a,davidchack:094710}
and non-equilibrium\cite{Bai05,Bai06,Baidakov10} molecular-dynamics
(MD) and metadynamics \cite{Angioletti10,Angioletti11} methods have
been developed for the calculation of interfacial free energies and
associated crystalline anisotropies. Similarly, both equilibrium and
non-equilibrium MD methods have been applied in calculations of the
magnitude and anisotropies of the mobilities of such interfaces.\cite{Asta09rev,Monk10,Hoyt2002}
To date, far less simulation work has been devoted to the investigation
of faceted crystal-melt interfaces.\cite{PhysRevE.78.031605,PhysRevE.80.050601,buta:074703,Henager09,PalafoxHernandez20113137,Yang20124960}
Additionally, the properties of steps at faceted crystal-melt interfaces
have been investigated primarily by kinetic Monte-Carlo simulations\cite{Beatty2000}
to date. To the best of the authors' knowledge, only one MD study
has been applied to the calculation of step mobilities,\cite{buta:074703}
and no direct MD calculations of step free energies have yet been
published. As a consequence, the molecular-level understanding of
the properties of steps at faceted solid-liquid interfaces remains
less advanced relative to those for rough crystal-melt interfaces
or steps at crystalline surfaces.

In this work we propose a method for the calculation of step free
energies $\gamma_{st}$ at faceted crystal-melt interfaces from equilibrium
MD simulations. The approach makes use of simulation geometries involving
coexisting solid and liquid phases, separated by a faceted solid-liquid
interface containing an island of solid or liquid on an otherwise
flat terrace. As described in the next section, the coexistence temperature
measured in such simulations is expected to vary with the curvature
(radius) of the islands, such that the relationship between radius
and the coexistence temperature measured in MD can be used to extract
orientation-averaged values of $\gamma_{st}$. The approach is illustrated
in this work in the application to steps in the Stillinger-Weber model
of Si.\cite{Stillinger85} The simulation methodology employed in
this application is described in Section \ref{sec:Methodology}, followed
by a presentation of results and a summary and discussion in Sections
\ref{sec:Results} and \ref{sec:Discussion}, respectively. In Section
\ref{sec:Results} we also present results illustrating that the Stillinger-Weber
model of Si shows a tendency to crystallize in the wurtzite, rather
than the diamond-cubic, structure at relatively low undercoolings.
While this finding is not directly relevant to the present study,
it is important for the detailed interpretation of simulations of
crystallization based on this potential.

\section{Theoretical Formalism\label{sec:Theoretical-Formalism}}

\begin{figure}
\includegraphics[width=0.8\textwidth]{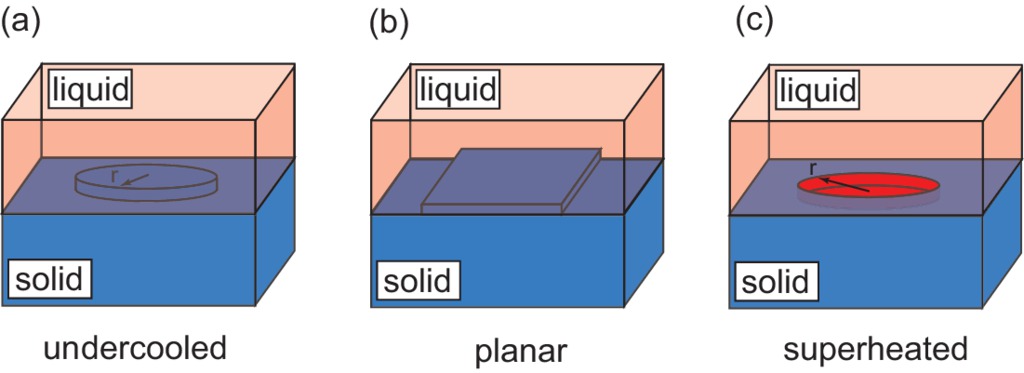}

\caption{Possible geometries of the solid-liquid interface with steps. a) Circular
solid island on liquid terrace. b) Equilibrium with planar geometry
of the step corresponds to the melting point $T_{m}$. c) Circular
liquid island on solid terrace. \label{fig:Schematics}}

\end{figure}

In Fig. \ref{fig:Schematics} we consider three possible geometries
of a faceted solid-liquid interface, where terraces of two different
{}``heights'' are present, separated by steps. In Figs \ref{fig:Schematics},a
and c the geometry corresponds to circular island of solid and liquid
phases, respectively. Figure \ref{fig:Schematics},b by contrast shows
a situation where terraces of different height are separated by straight
steps. When amounts of solid and liquid phases change due to melting
or crystallization the length of the step boundary changes in systems
with circular geometry, while it remains constant for the planar case.
Thus, equilibrium conditions for planar and circular geometries should
be different. 

We consider in more detail the equilibrium conditions for the systems
with a circular island, using the framework of classical nucleation
theory (CNT).\cite{Pimpinelli,YukioSaito} Consider a circular solid
island on a liquid terrace. The free energy of the system relative
to a state without an island is given by 

\begin{equation}
G=-\Delta\mu\rho_{A}\pi r^{2}+2\pi r\gamma_{st},\label{eq:G_of_a_step}
\end{equation}
where $\Delta\mu=\mu_{l}-\mu_{s}$ is the difference between the chemical
potential in the liquid and solid phases, $\rho_{A}$ is the atomic
density per unit area of the solid plane that forms the island, and
$r$ is the radius of the island. The size of the critical nucleus
$R$ corresponds to a maximum of $G(r)$ and is given by:

\begin{eqnarray}
R & = & \frac{\gamma_{st}T_{m}}{\rho_{A}H_{m}(T_{m}-T)},\label{eq:R_crit}
\end{eqnarray}
where we have used $\Delta\mu=\frac{H_{m}}{T_{m}}(T_{m}-T)$, with
$T_{m}$ corresponding to the equilibrium temperature for a planar
step geometry and $H_{m}$ is a heat of melting per atom.\cite{Pimpinelli,YukioSaito}
According to Eq. (\ref{eq:R_crit}) a system with a solid island of
radius $R$ is in equilibrium at temperature $T$ given as:

\begin{equation}
T=T_{m}(1-d/R)\label{eq:eq_temp_R}
\end{equation}
where the capillary length $d$ is given as:

\begin{equation}
d=\frac{\gamma_{st}}{\rho_{A}H_{m}}\label{eq:cap}
\end{equation}

Note that, from Eqs. (\ref{eq:eq_temp_R}) and (\ref{eq:cap}) a solid
island has a coexistence temperature below the bulk melting point.
By contrast a liquid island has a coexistence temperature above $T_{m}$
and corresponds to a superheated state. Finally, for the interface
geometry with a planar step shown in Fig \ref{fig:Schematics},b the
coexistence temperature is simply $T_{m}$, which follows from Eq.
(\ref{eq:eq_temp_R}) by taking the limit $R\rightarrow\infty$. Clearly,
in the latter case the equilibrium temperature is independent of island
size. The formalism discussed above neglects curvature corrections
to $\gamma_{st}$ as well as step interactions and stress effects.
However, these effects can in principle be incorporated by appropriate
generalizations of the thermodynamic formalism. 

The MD approach considered in this work makes use of the thermodynamic
formalism described in this section as follows. By performing equilibrium
MD simulations in the microcanonical ensemble (NVE), the constraint
of fixed total system energy can be used to stabilize the various
types of equilibrium states described above. By changing the system
energy, islands of different radii can be stabilized and the corresponding
system temperature derived. From the relationship between $R$ and
$T$, the step free energy can then be determined.

\section{Methodology of atomistic simulations\label{sec:Methodology}}

We demonstrate the approach outlined in the previous section by considering
the Stillinger-Weber model of elemental Si.\cite{Stillinger85} This
model represents one of the simplest classical interatomic potentials
that gives rise to faceted solid-liquid interfaces.\cite{PhysRevLett.56.155,PhysRevB.37.4647,PhysRevE.78.031605}
Several different values of the melting temperature have been reported
for this potential, from simulations with coexisting solids and liquids
separated by $(111)$ oriented interfaces, namely: $T_{m}=1677K$,\cite{buta:074703}
$T_{m}=1685K$\cite{PhysRevB.37.4647} and $T_{m}=1691K$.\cite{MorrisSi}
Unfortunately, it was not possible from these studies to know whether
the coexistence conditions corresponded to any of the different types
of interface geometries described in Fig. \ref{fig:Schematics}, or
to a flat terrace with no steps. We note that, in the latter case,
the coexistence temperature derived in an MD simulation can be different
from the true melting point, since the system can spend a significant
amount of time in such a state,\cite{Li11} due to the free energy
barrier required to nucleate islands of solid or liquid. During this
time thermal (uniform temperature) and mechanical equilibrium between
the phases can be established, but the phases are not in true equilibrium
with respect to phase change fluctuations, i.e. the chemical potentials
of the solid and liquid phases are not equal. To enable phase fluctuations
on the time scale of MD simulations it is desirable to have preexisting
steps at the interface. All equilibrium states implemented in this
study satisfy this requirement.

\subsection{Simulation block}

\begin{figure}
\includegraphics[width=0.8\textwidth]{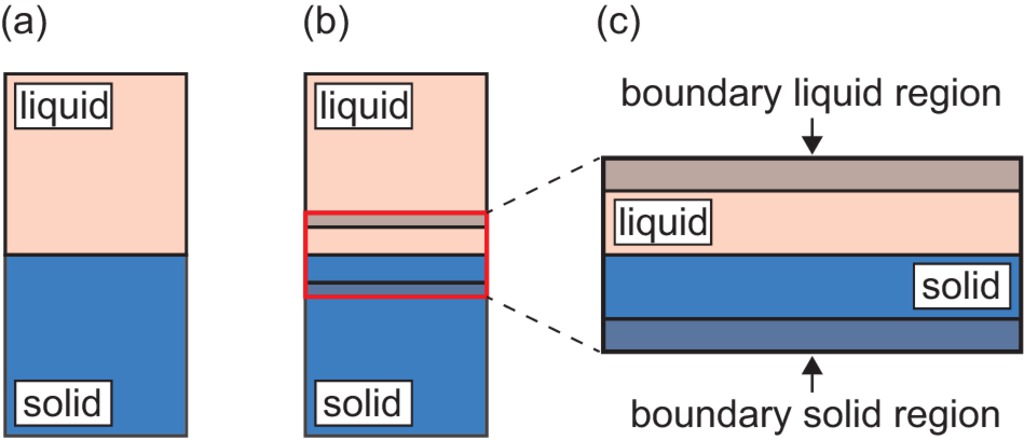}

\caption{Construction of the simulation block for modeling of an isolated interface.
a) Bulk solid and liquid at equilibrium. Dimensions of each phase
are approximately cubic. b) A region near interface is selected with
dimension normal the interface significantly smaller than the lateral
dimensions to create a new simulation block. c) A new simulation block
with two boundary regions. The solid boundary region is fixed in subsequent
simulations. The liquid boundary region moves as a rigid body to ensure
zero pressure in the liquid. \label{fig:Schematics-block}}
\end{figure}

To apply the simulation approach described above, we began by creating
a simulation block with diamond-cubic-structured solid and liquid
phases as shown schematically in Fig. \ref{fig:Schematics-block},a.
The (111)-oriented solid-liquid interface was perpendicular to the
$z$ direction of the block. The $x$ and $y$ directions were parallel
to crystallographic directions $[1\overline{1}0]$ and $[11\overline{2}]$
of the solid phase, respectively. The phases were equilibrated for
several nanoseconds in a microcanonical (NVE) ensemble. In the resulting
equilibrated state the dimensions of each phase were approximately
cubic. In the next step, we selected a region containing the solid-liquid
interface as well as regions of homogeneous solid and liquid phases
as shown in Fig. \ref{fig:Schematics-block},b. Atoms inside the top
and bottom layers were designated as boundary regions. The thickness
of each boundary layer was 10 Å. 

This procedure was used to create several simulation blocks with different
sizes. The number of atoms in the resulting systems ranged from 25196
to 226744. The dimensions $L_{x}$ and $L_{y}$ parallel to the interface
ranged from $5$ to $40$ nm. The dimension normal to the interface
$L_{z}=4.5$ nm was the same in all blocks. Periodic boundary conditions
were applied in the $x$ and $y$ directions. In the $z$ directions
the boundary conditions were not periodic. The positions of atoms
inside the solid boundary region were fixed during the simulations.
The top boundary region of liquid atoms moved as a rigid body during
the simulations to ensure zero pressure inside the liquid phase below
the boundary region.

Such boundary conditions were chosen for two reasons. First, to apply
the formalism described in the previous section it is necessary to
model only one solid-liquid interface, which is impossible with periodic
boundary conditions in the $z$ direction. Second, the rigid body
boundary condition was chosen over the open liquid surface to isolate
the solid-liquid interface from the effect of surface capillary waves
and density variations. Thus, the rigid slab of liquid (and solid)
mimics an infinitely large homogeneous phase. Due to the use of relatively
small lengths in $z$ the values of the step energies computed below
could be affected by finite-size effects, because we are constraining
the phonon spectrum in the crystal and the density oscillations in
the liquid near the solid-liquid interface. We emphasize however that
the primary purpose of the present work is to demonstrate the adiabatic-trapping
methodology for computing step free energies. Finite size effects
could be straightforwardly investigated in future studies focused
on precise values for specific systems.

\subsection{MD simulations}

Molecular dynamics simulations were performed using the Large-scale
Atomic/Molecular Massively Parallel Simulator (LAMMPS) software package.\cite{Plimpton19951}
Thermal expansion of solid Si at $T=1679$ K was computed from a simulation
in an NPT ensemble with zero pressure. Constant temperature and pressure
in the simulation were imposed using a Nosé-Hoover thermostat\cite{Melchionna93}
and an Anderson barostat,\cite{Andersen1980} respectively.

Molecular Dynamic simulations of the solid-liquid interface were performed
in a microcanonical ensemble (NVE) for times up to 50 ns after 5ns
equilibration stage, using a time step of $1$ fs. During the simulations
$x$ and $y$ dimensions of the simulation block were kept fixed,
with the lattice parameter in the solid part of the block equal to
the stress free lattice parameter at $T=1679$ K. 

During the simulations snapshots of the system were saved every 0.5
ns. The snapshots contained positions of atoms, energies and stresses
and were used for post processing. The equilibrium temperature $T$
of the simulation was calculated from an average of the kinetic energy
over the production stage.

\subsubsection{Trapping procedure}

In canonical simulations (NVT) solid/liquid islands on liquid/solid
terraces are usually not in equilibrium with the rest of the system:
subcritical islands shrink, whereas supercritical islands grow to
completely cover the terrace. Critical islands correspond to a state
of unstable equilibrium. To compute sizes of critical nuclei we employed
a trapping procedure used previously in a study of heterogeneous nucleation
at grain boundaries.\cite{Frolov11a} In a finite system and a restricted
ensemble (NVE) equilibrium between the island and the rest of the
system can be stabilized due to the constraint of constant energy.
To achieve a state of a stable equilibrium, fluctuations in the size
of the island must produce sufficient changes in temperature of the
system. As a consequence the $z$ dimension of the simulation block
cannot be too large. Also lateral dimensions of the island have to
be comparable to $x$ and $y$ dimensions of the simulation block.
For the reasons above, for each size of the simulation block a structure
with an island of a size within a certain range had to be carefully
prepared. If the size of an island was too small, it could completely
disappear by a fluctuation during a 50 ns long simulation. Additionally,
a very large island could hit the boundary conditions and change its
morphology from circular to planar, thus changing the nature of the
equilibrium state.

To prepare a simulation block with a suitable size of an island, the
radius of the solid/liquid island on the terrace can be modified by
injecting/removing heat into a system. This was achieved by modifying
velocities of atoms in the input file and equilibrating the system
during a several nanosecond run in the NVE ensemble. We prepared simulation
blocks with solid/liquid islands embedded inside the liquid/solid
terrace as well as blocks with solid and liquid terraces separated
by a planar step boundary. The former states correspond to curved
equilibrium with equilibrium temperature depending on the size $R$
of the islands, while the later case corresponds to planar equilibrium
with temperature independent of the sizes of the terraces.

\subsubsection{Data analysis}

To analyze the structure of the interface and calculate the sizes
of islands it was necessary to identify atoms belonging to solid or
liquid phases. This was done using a structural order parameter employed
in the previous studies of the Stillinger-Weber Si system.\cite{buta:074703,PhysRevE.78.031605,PhysRevE.80.050601}
The calculated order parameter takes values close to 1 inside solid
phase and 0 inside liquid phase. A threshold value of the order parameter
was calculated as an average of the values inside solid and liquid
phases. Atoms with the value of the order parameter higher than this
threshold value were identified as solid, while all other atoms were
identified as liquid. 

The structure of the solid-liquid interface was analyzed by identifying
isolated solid/liquid islands on the liquid/solid terrace. For each
atom belonging to solid/liquid phase a list of the nearest neighboring
atoms of the same phase were constructed. Then a random walk through
neighboring atoms was implemented in the following way. First, a random
atom of a given phase was selected and assigned a cluster number.
Second, its neighbor belonging to the same phase was randomly selected
and assigned the same cluster number. Thus, repeating the procedure
all atoms belonging to one cluster can be identified. The procedure
was previously employed to identify clusters in a binary liquid.\cite{Frolov10prl}
The largest island identified represents the nucleus, while all other
islands are fluctuations containing only a few atoms. These small
islands are present in the structure of the terrace at a given temperature.\cite{PhysRevE.78.031605}

Our cluster analysis of the interface allows us to calculate the center
of mass of the largest island on the terrace. To compute an average
size and shape of the island we average the atomic density of the
instantaneous islands relative to instantaneous centers of mass. Once
the average atomic density is computed, the radius $R$ of an island
is calculated as a distance from the center of mass to the point where
the average density value is equal half of the density of solid (111)
layer.

\section{Results\label{sec:Results} }

\subsection{Interface Geometries}

\begin{figure}
\begin{centering}
\includegraphics[width=1\textwidth]{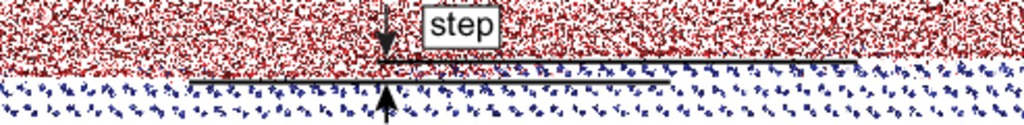}
\par\end{centering}

\caption{(111) solid-liquid interface with a step. Solid atoms are colored
in blue while liquid atoms are shown in red. The coloring is according
to the order parameter described in the main text. The upper (111)
plane is incomplete and forms a step with the liquid terrace on the
right. The image was produced with the ATOMEYE visualization program.\cite{Li03}
\label{fig:step}}
\end{figure}

Fig. \ref{fig:step} shows a cross-sectional view of a region of a
$(111)$ solid-liquid interface containing a step separating two terraces
of different heights. Atoms belonging to the liquid phase are colored
in red, while solid atoms appear in blue. The $[1\overline{1}0]$
crystallographic direction in the solid phase is normal to the plane
of the figure. The top solid $(111)$ plane is incomplete and forms
a step boundary with the liquid terrace. The height of the step is
equal to the bilayer spacing of the diamond-cubic structure along
the $[111]$ direction.

\begin{figure}
\includegraphics[height=0.7\textheight]{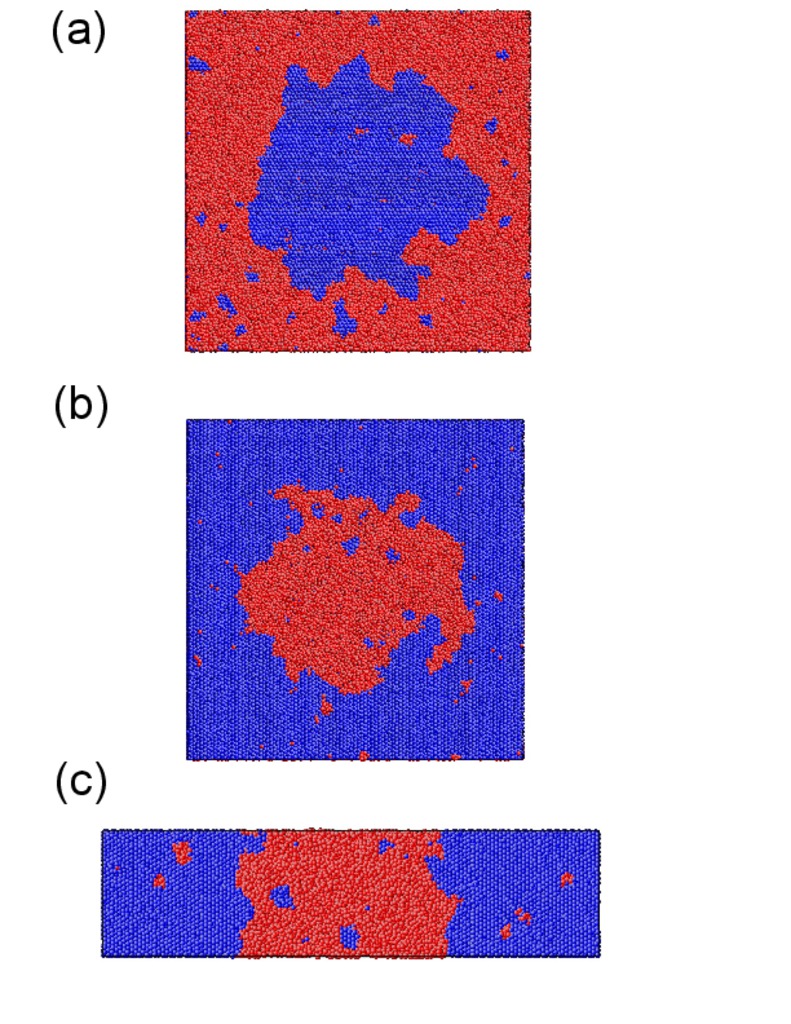}

\caption{Structures of the solid liquid-interfaces with different types of
terrace geometries. (a) A snapshot of a solid island inside a liquid
terrace at $T\sim1678$ K (undercooling). (b) Snapshot of a liquid
island inside a solid terrace at $T=1685.4\pm0.7$ K (superheating).
(a) and (b) correspond to curved equilibrium. (c) Solid and liquid
terraces separated by a planar step boundary at $T_{m}=1682.0\pm1.0$
K. The images were produced with the ATOMEYE visualization program.\cite{Li03}\label{fig:nuclei}}
\end{figure}

Fig. \ref{fig:nuclei} shows plan views of snapshots from the equilibrium
configurations of faceted (111) solid-liquid interfaces with different
island geometries. Figs \ref{fig:nuclei},a and b represent curved
equilibrium states of a solid island on a liquid terrace, and liquid
island on a solid terrace, respectively. Both simulation blocks have
the same size with lateral dimensions of $30\times30$ nm. These snapshots
illustrate that the instantaneous structure of the step is extremely
rough, with the shape of the island being quite complicated and characterized
by \textquotedblleft{}overhangs\textquotedblright{} and large deviations
from a perfect circular geometry. The equilibrium temperatures calculated
for these solid and liquid islands were $T\sim1678$ K and $T=1685.4\pm0.7$
K, respectively. Thus, for the same size of the simulation block the
calculated equilibrium temperature was not the same due to the difference
in geometry of the atomic layer at the solid-liquid interface.

Fig. \ref{fig:nuclei} c shows a plan-view snapshot illustrating the
equilibrium state of the interface with a planar geometry of the steps
at the calculated temperature $T=1682.4\pm0.6$ K. The $z$ dimension
normal to the solid-liquid interface was the same as in all other
blocks studied. The lateral dimensions were $10\times40$ nm$^{2}$
parallel to the $x$ and $y$ axis respectively. The simulation block
contains two step boundaries. This geometry with one dimension of
the simulation block smaller than the other is more advantages for
simulations of planar-step equilibrium. It ensures that the liquid
or solid part of the interface layer does not transform into an isolated
island with circular boundaries during a long simulation. Simulations
with planar step geometries were also performed for systems with lateral
periodic dimensions of $5\times10$ nm$^{2}$ and $10\times30$ nm$^{2}$.
Identical equilibrium temperatures were calculated for these blocks
within the statistical accuracy of the simulations, as discussed in
further detail below.

\subsubsection{Wurtzite Formation}

\begin{figure}
\includegraphics[width=0.7\textwidth]{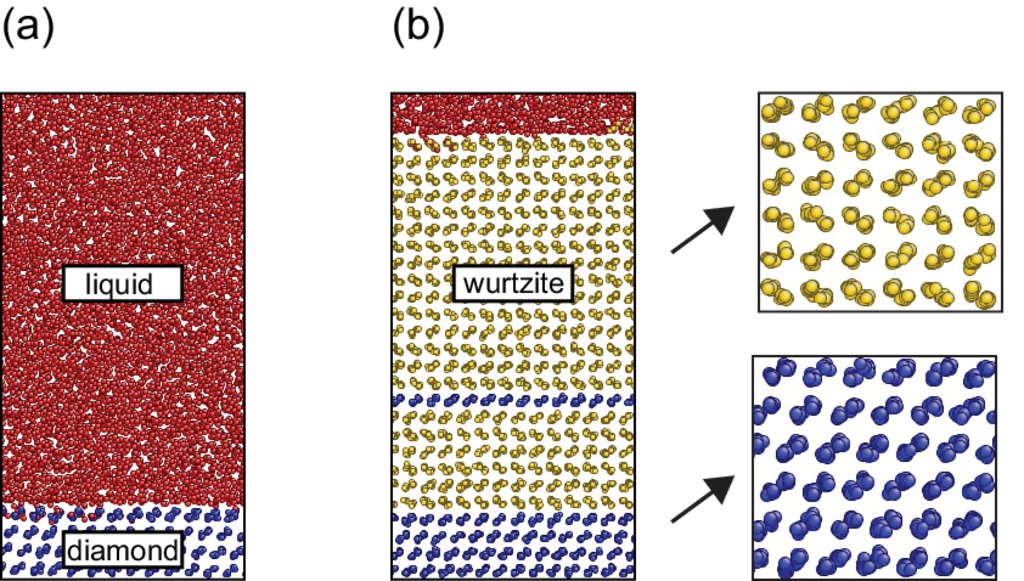}

\caption{Simulations of solidification at $T=1670$ K. Initially the simulation
block contains liquid and a crystalline solid in the diamond structure.
The liquid phase in a) solidifies primarily into wurtzite structure
in b). The images were produced with the ATOMEYE visualization program.\cite{Li03}\label{fig:Solodi}}

\end{figure}

In the simulations with solid circular islands in a liquid terrace
we observed nucleation of new islands showing a registry with the
underlying solid layer corresponding to the wurtzite, rather than
the diamond cubic structure. These new {}``wurtzite'' islands were
observed to grow and consume the preexisting {}``diamond'' islands.
After this transformation the islands were observed to remain in the
wurtzite state. Previous MD simulations of solidification with $(111)$
oriented solid-liquid interfaces\cite{PhysRevB.37.4647} and of isolated
solid nuclei\cite{Beaucage05} reported that the Stillinger-Weber
Si potential solidifies in a random mixture of stacking sequences.
These simulations modeled large undercoolings, and the observations
could be a consequence of rapid solidification kinetics. To investigate
this issue further, we performed a separate simulation of solidification
at 1670 K (\textasciitilde{}12K undercooling) in the NP$_{z}$T ensemble,
featuring fully periodic boundary conditions with fixed area parallel
to the interface and dynamic periodic lengths normal to the interface
to ensure zero normal stress. The simulation block had dimensions
$5\times5\times18$ nm$^{3}$, with two (111) solid-liquid interfaces
normal to the $z$ direction. Figs. \ref{fig:Solodi},a and b show
the initial state of the simulation block with the solid phase in
the diamond structure and the simulation block after 110 ns long simulation,
respectively. The liquid phase crystallized into wurtzite phase. Only
one out of 32 newly grown (111) layers was observed to have the diamond-cubic
stacking sequence.

The observation of the nucleation and growth of the wurtzite structure
during simulations at low undercoolings is likely an artifact of the
Stillinger-Weber potential. Specifically, due to its short cutoff
distance the potential gives identical zero-temperature energies for
the diamond and wurtzite structures.\cite{Beaucage05} The wurtzite
phase may appear during solidification for two reasons. First, it
may be more stable than the diamond phase at high temperatures, which
means that it would have a higher melting point. Second, the liquid
may crystallize into wurtzite due to purely kinetic effects. For example,
if the step free energy of a wurtzite island is lower than that of
a diamond island, a monolayer of wurtzite will nucleate first. At
high undercoolings nucleation barriers for diamond and wurtzite structures
become nearly zero and the nucleated structure is determined by random
fluctuations. As a result, it is understandable that at large undercoolings
random sequence of these two phases can be observed.\cite{PhysRevB.37.4647}
The melting point of the wurtzite phase was calculated using simulation
blocks with planar geometries for the steps and was found to be $T_{M}^{W}=1681.5\pm1.3$
K, which is nearly identical with that of the diamond phase. Therefore,
the second scenario, involving kinetic stabilization of the wurtzite
phase, is more consistent with our simulation results.

\subsection{Island Radius versus Temperature and Step Free Energy}

Because of the diamond to wurtzite transformation we were unable to
calculate the relationship between island radius ($R$) and temperature
($T$) for diamond-cubic solid islands on the liquid terrace. Accurate
equilibrium temperature calculations require tens of nano-seconds
of sampling time and we were unable to accumulate the necessary statistics
before transformations to the wurtzite structure occurred. In the
simulations with liquid islands on solid terrace, nucleation of wurtzite
islands is unfavorable because the temperature of the simulation is
higher then the meting point of the wurtzite phase. As a result we
were able to model liquid islands for any desired period of time to
compute the relationship between $R$ and $T$.

\begin{figure}
\begin{centering}
\includegraphics[width=0.8\textwidth]{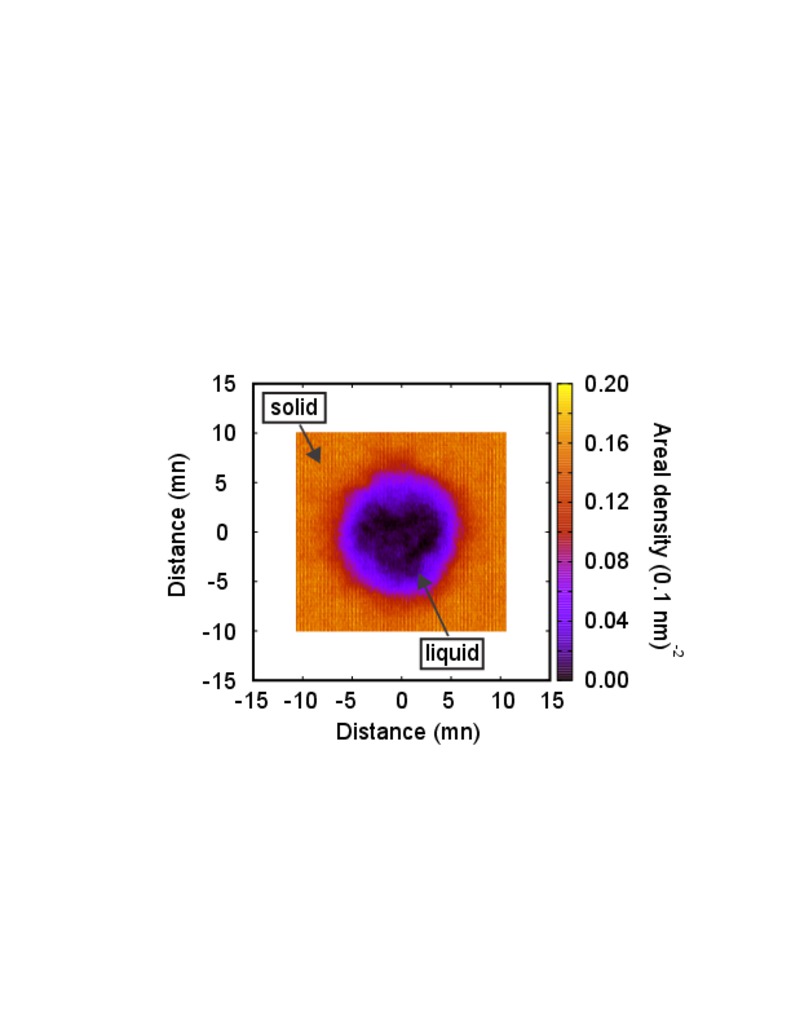}
\par\end{centering}

\caption{Average density of atoms identified as solid according to the order
parameter at $T_{m}=1685.4\pm0.7$ K obtained by averaging of instantaneous
snapshots. The orange region corresponds to a solid terrace with density
close to that of a perfect $(111)$ plane ($\rho_{A}=0.157\times10^{-20}$
m$^{-2}$). The near zero density of solid atoms in the dark middle
region of the plot represents the liquid nucleus. This density map
was used to compute the size of the critical nucleus $R$.\label{fig:PFpicture}}
\end{figure}

Fig. \ref{fig:PFpicture} demonstrates an average equilibrium shape
of a liquid island on solid terrace at $T=1685.4\pm0.7$ K. The shading
in the image corresponds to the average density of solid atoms, as
indicated in the legend. The solid-atom density inside the island
is nearly zero, while it is close to the areal density of a (111)
solid plane away from the island. The thickness of the transition
region where the density changes between these limiting values is
about 5 nm, which is comparable with the radius of the island in Fig.
\ref{fig:PFpicture}. While the instantaneous shapes of the island
are very complicated (c.f., Fig. \ref{fig:nuclei} a and b) the average
shape shown in Fig. \ref{fig:PFpicture} appears to be almost perfectly
circular. Roughness of the step as well as instantaneous deviations
from the circular shape and size fluctuations all contribute to the
broadening of this region. While the first two processes are most
likely to be the dominant ones, the third process is always present
and strongly affected by the size of the block and the island.

\begin{figure}
\begin{centering}
\includegraphics[width=0.7\textwidth]{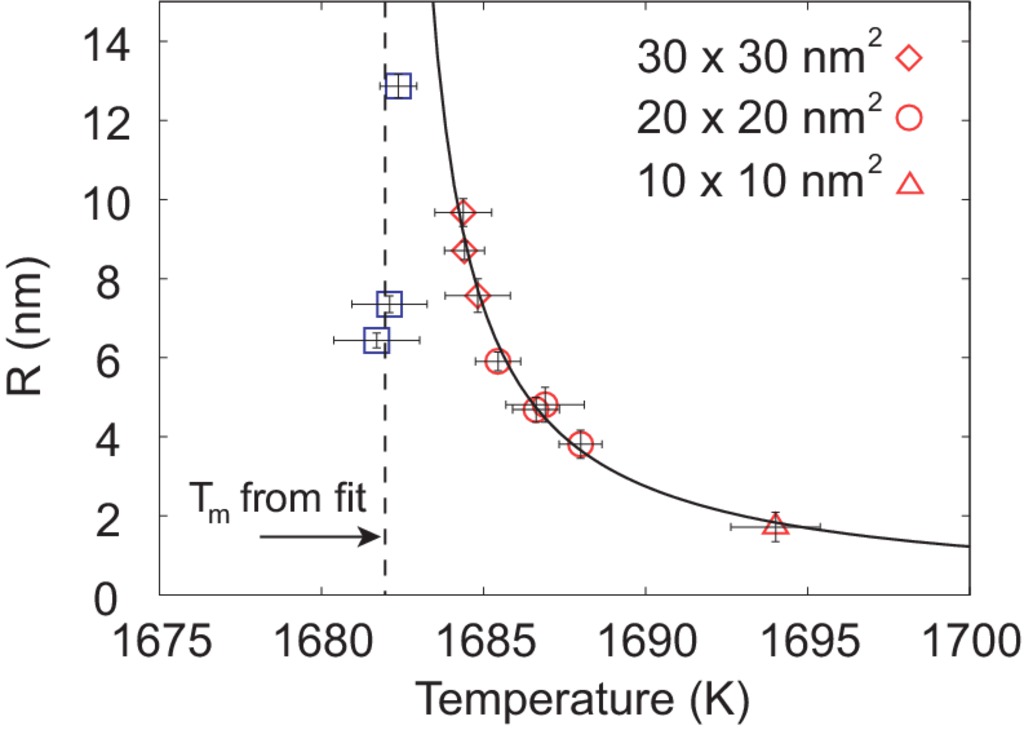}
\par\end{centering}

\caption{Radius of critical nucleus ($ $ \textifsymbol[ifgeo]{54}, \Circle{},
\textifsymbol[ifgeo]{49} ) as a function of temperature. The dimensions
of the simulation blocks parallel to the solid-liquid interface are
indicated on the plot. The melting point for planar step geometries
was calculated for different sizes of the simulation block ( $\square$
) . The continuous line was fit to discrete data points using Eq.
(\ref{eq:R_crit}). The dashed line indicates the melting point obtained
from fitting the curved equilibrium points with Eq. (\ref{eq:R_crit}).
 \label{fig:RvsT}}
\end{figure}

Fig. \ref{fig:RvsT} summarizes calculations of the equilibrium island
radius and corresponding equilibrium temperatures. The discrete points
on the figure correspond to individual MD simulations. Squares correspond
to simulations having steps with planar geometry (e.g., Fig. \ref{fig:nuclei},c),
while other symbols represent simulations of isolated liquid islands
on a solid terrace (e.g., the geometry shown in Figs. \ref{fig:nuclei},b
and \ref{fig:PFpicture}). The continuous lines on the plot were obtained
by fitting the curved equilibrium data points with Eq. (\ref{eq:R_crit})
using $\gamma_{st}$ and $T_{m}$ as fitting parameters. The data
points with planar step geometry were not included in the fitting.
The excellent agreement between the fit and the discrete data points
allows us to conclude that equilibrium island radius $R$ is inversely
proportional to $\Delta T=T-T_{m}$. The melting point obtained from
the fit was $T_{m}=1681.96\pm0.05$ K, which is identical within statistical
uncertainties with the equilibrium melting temperature calculated
directly from MD simulations with planar step geometries: $T_{planar}^{MD}=1682.0\pm1.0$
K, computed as an average over the three independent simulations with
different sizes.

Using the data for latent heat of melting for Stillinger-Weber Si,
$H_{m}=31$ kJ/mole ($5\times10^{-20}$ J/atom),\cite{MorrisSi} the
areal atomic density of the (111) plane $\rho_{A}=0.157\times10^{20}$
m$^{-2}$, and the value of the melting point given in the previous
paragraph, we calculate from the fit shown in Fig. \ref{fig:RvsT}
a value for the step free energy of $\gamma_{st}=0.103\pm0.005\times10^{-10}$
J/m, where the error bar was determined from the fitting. Dividing
this value by the (111) interplanar distance $d^{(111)}=3.14$ Å,
we obtain a value for the so-called perimeter free energy\cite{Mullins2000}
of $\gamma_{st}^{p}=0.033\pm0.0016$ J/m $^{2}$. The perimeter free
energy has units of energy per unit area and can be compared with
solid-liquid interface free energy.

The solid-liquid interface free energies for Stillinger-Weber Si calculated
using the cleaving technique\cite{Apte08} for (100), (111) and (110)
orientations are $0.42\pm0.02$ , $0.34\pm0.02$ and $0.35\pm0.03$
J/m$^{2}$ respectively. These values are approximately an order of
magnitude larger than the perimeter free energy quoted above. Similarly,
multiplying the interface free energy by the thickness of one (111)
monolayer would give an estimate for $\gamma_{st}$ of $1.16\times10^{10}$
J/m, which is an order of magnitude larger than the step free energy
calculated by MD in the present work.

\section{Discussion and conclusions\label{sec:Discussion}}

In this work we presented a method for MD calculations of the free
energies of steps of faceted solid-liquid interfaces. The approach
makes use of simulations in which the faceted interface contains islands
with curved steps of differing radii. For a range of island radii
the coexistence temperature is derived from equilibrium MD simulations
and the resulting relationship between $R$ and $T-T_{m}$ is used
to back out $\gamma_{st}$.

The approach was demonstrated in an application to (111) faceted interfaces
in the Stillinger-Weber model of elemental Si. We simulated interfaces
with planar geometries of the steps as well as shapes of islands that
were circular on average. In the later case, there were two different
states, corresponding to a solid island in a liquid terrace, or a
liquid island in a solid terrace. We demonstrated that the calculated
equilibrium temperature depends on structure of the interface. When
the step separating solid and liquid parts of the interface is planar,
the calculated equilibrium temperature $T_{m}=1682.0\pm1.0$ K was
found to be remarkably consistent and independent of the size of the
islands. In the case of equilibrium with a circular island geometry,
the island radius $R$ was found to be inversely proportional to $T-T_{m}$.

The island size dependence on equilibrium temperature allowed us to
compute step free energy $\gamma_{st}$. The calculated step free
energy is an order of magnitude smaller than the product of interface
free energy and thickness of a (111) monolayer. The difference between
the two quantities is not surprising. When a faceted interface undergoes
a roughening transition, the step free energy becomes zero, while
solid-liquid interface free energy generally remains finite. Therefore,
these two properties are not directly related. Large island shape
fluctuations observed in MD and the roughness of the step structures
correlate with small value of the step free energy, and indicates
that the interface may not be far from a roughening transition.

A relatively wide range of equilibrium temperatures have been reported
in the literature for coexistence simulations with (111) oriented
interface. Our simulations provide an insight why, in general, this
should be the case. The equilibrium temperatures calculated in this
work ranged from 1682 K to 1695 K, depending on the exact nature of
the interface geometry. This range would actually be twice large,
if we include the undercooled branch where quantitative analysis of
the data was not possible due to the formation of layers with wurtzite
stacking. The current results thus illustrate that variations of equilibrium
temperature within 26 K range are possible when simulating solid-liquid
systems separated by the faceted (111) interface. During solid-liquid
coexistence simulations with (111) interfaces with large bulk phases,
the interface geometry may transition by fluctuations between undercooled,
planar or superheated states. The interface can also adopt a geometry
when the interface layer is complete (state with no step) and spend
a significant amount of time trapped in this metastable state. As
a result, depending on the initial conditions, amounts of phases and
time of calculations, the resulting average temperature may deviate
significantly from the true melting temperature. This study shows
that for a consistent calculation of $T_{m}$ using the coexistence
approach, for systems with faceted interfaces, an interface with a
planar geometry of a step is desirable.

The approach of modeling of plane and curved equilibrium islands to
extract values of step free energies can be extended to multicomponent
systems. In the later case MD simulations in an NVT ensemble or constant
composition Monte Carlo\cite{Stukowski2012} simulations are more
appropriate. The stability of the island is maintained by constant
average composition in the system, rather than a constant energy constraint.
Performing a series of isothermal calculations at different temperatures,
one could recover the dependence of the step free energy on composition/temperature.
This dependence is an important ingredient in understanding and mesoscale
modeling of layer-by-layer growth in the context of important technological
processes such as nanowire growth by the vapor-liquid-solid mechanism.
For example, in applications of such an approach to Si in contact
with a eutectic Au-Si melt, the equilibrium temperatures probed in
the simulations would be significantly lower than melting point of
elemental Si. It is possible, that the step structure will not be
as rough as than observed in the present simulation. Specifically,
it is possible that the steps would become significantly straighter,
such that the islands would develop {}``polygonal'' shapes, using
the terminology from Ref. $ $\cite{ChernovMRS}. Investigation of
equilibrium shape of such islands and their evolution during growth
would be of fundamental interest.

Finally, this work points to an important shortcoming of the Stillinger-Weber
potential in modeling crystal growth in Si. Due to the small radius
of interaction, the potential does not differentiate between the diamond
and the wurtzite structures, and the energies of these structures
are identical at zero temperature.\cite{Beaucage05} In the present
simulations we observed that at low undercoolings the liquid crystallizes
into the wurtzite structure, and interfacial islands with diamond
stacking are unstable with respect to growth of islands with wurtzite
stacking. The appearance of the wurtzite phase is often observed during
nano-wire growth\cite{PhysRevLett.99.146101} and can be due to a
difference in step free energies. It is possible that at different
undercoolings the liquid would crystallize into phases are that are
metastable in the bulk for kinetic reasons. To explore this issue
further for growth of Si crystals would be of great interest, but
for such purposes it would be desirable to employ a potential that
better describes the relative energetics of the competing diamond
and wurtzite phases.
\begin{acknowledgments}
This research was supported by the US National Science Foundation
under Grant No. DMR-1105409. Use was made of computational resources
provided under the Extreme Science and Engineering Discovery Environment
(XSEDE), which is supported by National Science Foundation grant number
OCI-1053575. T.F. was also funded by the Miller Institute. 
\end{acknowledgments}

\bigskip{}

\clearpage{}

\clearpage{} 
\end{document}